\documentclass[review,number,sort&compress]{elsarticle}

\usepackage{amssymb}
\usepackage{amsmath}
\usepackage{gensymb}

\usepackage{wasysym}

\usepackage{graphics} 
\usepackage{graphicx}
\usepackage{color}
\usepackage{subfigure}

\usepackage{rotating}
\usepackage{textcomp}
\usepackage{subfigure}
\usepackage{xcolor}

\usepackage{amsmath,amsthm,amssymb,amsfonts}
\usepackage{units}
\usepackage{hyperref}

\usepackage{changepage}

\usepackage{dcolumn}

\usepackage{lineno}

\usepackage{verbatim}
\usepackage{here}

\begin{document}
\begin{frontmatter}
\title{
A prestorage method to measure
neutron transmission of ultracold neutron guides}
%
%
%
%
\author[PSI]{B.~Blau}
\author[PSI]{M.~Daum}
\author[PSI,ETH]{M.~Fertl\fnref{MF}}

\author[ILL]{P.~Geltenbort}
\author[PSI,ETH]{L.~G\"oltl}
\author[PSI]{R.~Henneck}
\author[PSI,ETH]{K.~Kirch}
\author[PSI]{A.~Knecht}
\author[PSI]{B.~Lauss\corref{cor1}}
\author[PSI]{P.~Schmidt-Wellenburg}
\author[PSI]{G.~Zsigmond}


%
\address[PSI]
{Paul Scherrer Institute (PSI), CH-5232 Villigen PSI, Switzerland}
\address[ETH]
{Institute for Particle Physics, Eidgen\"ossische Technische Hochschule (ETH), Z\"urich,  Switzerland}
\address[ILL]
{Institute Laue-Langevin (ILL), 71 avenue des Martyrs, F-38000 Grenoble, France }

\cortext[cor1]{Corresponding author, bernhard.lauss@psi.ch}

\fntext[MF]{Now at University of Washington, Seattle WA, USA.}


\begin{abstract}
There are worldwide efforts
to search for physics beyond the Standard Model of particle physics. 
Precision experiments using ultracold neutrons (UCN)
require very high intensities of UCN.
Efficient transport of UCN
from the production volume to the experiment is therefore of great importance. 
We have developed a method using prestored UCN
in order to quantify UCN transmission in tubular guides.
This method simulates the final installation at
the Paul Scherrer Institute's UCN source where 
neutrons are stored in an intermediate storage vessel
serving three experimental ports.
This method allowed us to qualify UCN guides for their intended use
and compare their properties.
\end{abstract}



\end{frontmatter}






\section{Introduction}
\label{intro}

Ultracold neutrons (UCN) are defined via
their unique property of being 
reflected under any angle of incidence from the surface of suitable materials
with high material optical 
potential $V_F$ (Fermi potential) \cite{Golub1991,Ignatovich1990}.
This is only occurring 
for neutrons with very low kinetic energies in the \,neV range,
corresponding to velocities below 7\,m/s or 
temperatures below 3\,mK.
Hence their name ''ultracold''.
Well-suited materials are e.g. stainless steel, Be, Ni, NiMo alloys or diamond-like carbon 
which display total UCN reflection up to their respective 
$V_F$ of 190, 252, 220 or 210 - 290\,neV, respectively \cite{Golub1991,Ignatovich1990}.
Closed containers of such materials can confine UCN and serve as UCN storage vessels.
Evacuated 
tubes or rectangular shaped guides made of or coated with 
materials of high $V_F$
can be used to transport UCN over distances
of several meters.

The ultracold neutron source at PSI 
\cite{Anghel2009,Lauss2011,Lauss2012,Lauss2014,Goeltl2012}
is now in normal operation.
About ten meters of neutron guides
are necessary to transport UCN from the 
intermediate storage vessel to one of the three beam ports,
traversing the several meter thick biological shield.
The UCN guides, made from coated glass or coated stainless steel,
are housed inside a stainless steel vacuum system.
High vacuum conditions are required
in order not to affect the UCN storage and transport properties.

The main thrust for the construction and operation of 
high intensity UCN sources \cite{Kirch2010} comes from the needs of
high precision experiments like the search for a 
permanent electric dipole moment of the neutron \cite{Baker2011,Afach2014,Afach2015Exotic}.
Efficient transport of UCN from production to the experiment is
a necessity.
In addition, installation of the guides is a complex and lengthy procedure. 
Moreover, the replacement of a guide would cause a long shutdown. 
Therefore, it was decided to install only UCN guides
tested with UCN and with known good UCN transmission.
We have developed the 
prestorage method, 
described in section~\ref{sec:prestorage},
which allowed us to perform a quality check on UCN guides
and a quantification of the UCN transmission properties
before final installation.

\subsection{UCN transmission}    

Ignatovich \cite{Ignatovich1990} dedicates a long chapter in his book 
to ``Transporting UCN'' and gives the definition of 
UCN transmission as
``the transmission of a neutron guide is the ratio of UCN flux at the output to the flux at the input''.
This is applicable when regarding continuous UCN sources
and continuous experiment operation.
%
In storage type experiments, 
when an experimental chamber has to be filled
within a given time period and using a UCN source with
UCN intensity decreasing with time, the integrated 
number of UCN and the UCN passage time is relevant.
In our method 
the transmission of the time-integrated counts 
will be studied instead of the flux transmission. 

\begin{figure}
\centering
\includegraphics[width=0.75\textwidth]{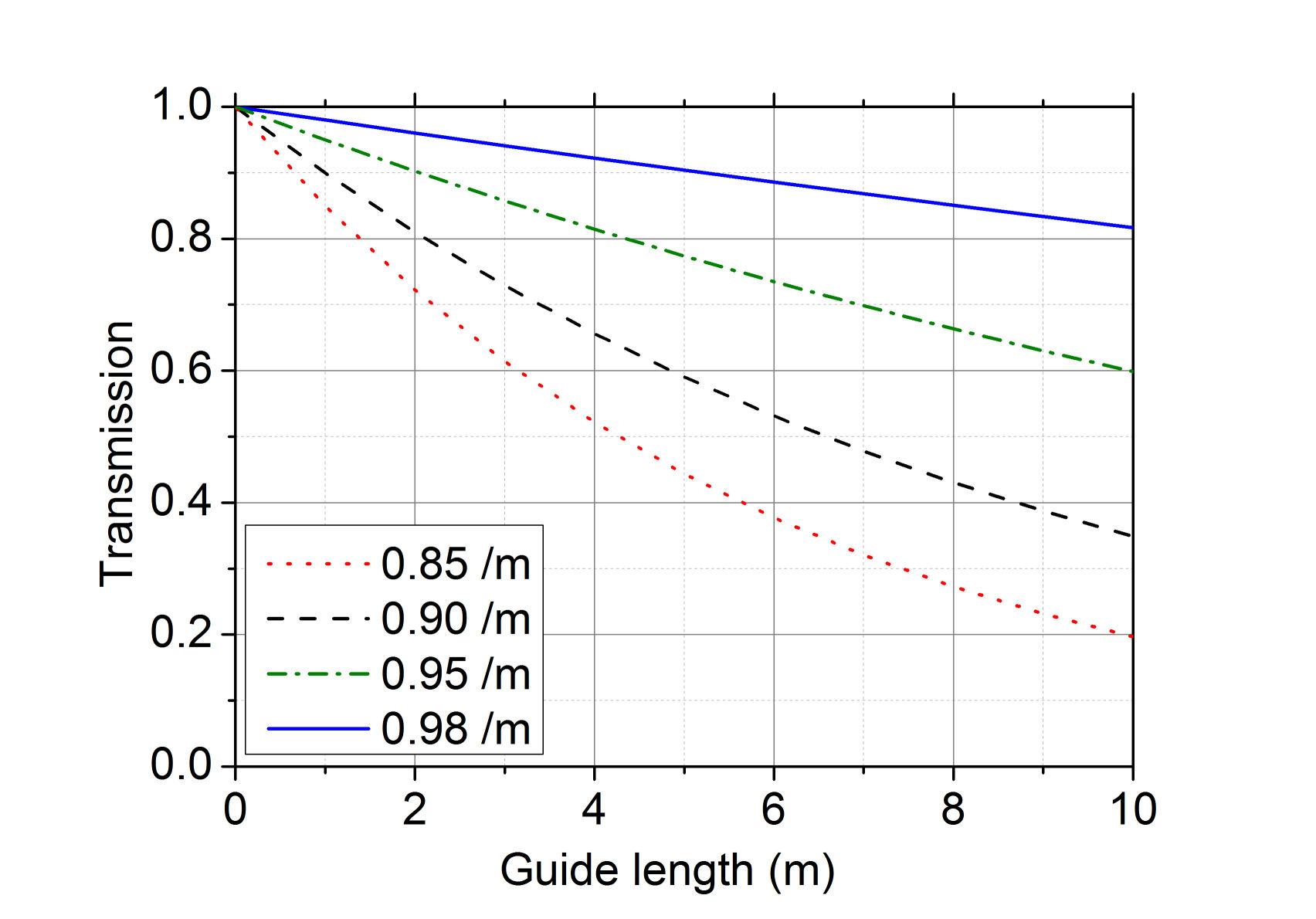}
\caption{
Calculated transmission probability of UCN 
through a guide with increasing length
for
different normalized transmission probabilities between 
0.85\,/m and 0.98\,/m}
\label{transmission-simulation}
\end{figure}

%
Independent of the materials used, 
UCN transmission
increases with guide diameter and
decreases with  
guide length L.
For comparison 
between different types of guides
the
normalized transmission per meter
(T$_{\rm norm}$) is used.
The total guide transmission T then 
follows as:
\begin{equation}
T~=~(T_{\rm norm})^\frac{L}{1 [m]}.
\end{equation}
%
Fig.~\ref{transmission-simulation}
shows the calculated behavior of the UCN transmission
with increasing guide length
for normalized UCN transmissions
in a realistic range for good UCN guides.
A length of about 10\,m is necessary at the PSI source 
to pass the biological shielding.
Fig.~\ref{transmission-simulation} demonstrates 
the importance of even small improvements in UCN transmission 
for such long installations.
In general, the properties of the materials
and UCN exposed surfaces, 
are decisively separating the good and the bad guides.

In the past, 
various measurements have been made to define and measure
the properties of UCN guides with early attempts
summarized in \cite{Ignatovich1990}.
The topic is also treated in recent publications 
\cite{Nesvizhesky2006,Plonka2007,Altarev2007,Frei2010,Daum2014},
but experiments have notoriously been difficult
with results not-necessarily transferable to other measurements.
Problematic issues were necessary assumptions on neutron flux,
neutron energy distribution, detector efficiencies, and reproducibility
of installations concerning 
e.g. small gaps in the setup causing UCN losses.


Assuming a single number for UCN transmission is a simplification 
as the transmission probability depends on the
kinetic energy and angular distribution of the neutrons.
Furthermore, it is important to know how fast UCN traverse the
guide, i.e.\ how fast one can fill an experiment on the exit side,
which directly correlates to specular reflectivity
and integral transmission.
In principle it
would be possible to repeat our measurements with monoenergetic UCN
with a more complicated setup.

The main parameters which define the 
neutron transmission of a UCN guide 
can be summarized as follows:

\begin{itemize}

\item{Surface roughness: 
The total reflections of neutrons
from surfaces can be classified in specular reflections,
where the angle of incidence 
is co-planar and 
equals the reflection angle and
diffuse reflections, 
where the reflection angle is independent of the incident angle and
follows a cosine distribution with 
respect to the perpendicular direction \cite{Golub1991}. 
This simple view is valid for roughness values much larger than the neutron wavelength. 
For very low roughness, e.g. for
highly polished copper 
or coated glass surfaces \cite{Steyerl1972},
diffraction effects become important and the probability of diffuse 
reflections will depend on the incident angle and neutron elocity (see also \cite{Ignatovich1990}).
The influence of surface roughness on UCN reflection has been studied 
recently in detail using flat plates as reflector \cite{Heule2010}.
%
%
High UCN transmission 
is obtained with negligible diffuse reflections.
which results
in a short passage times for the UCN through the guide.
Hence, low surface roughness is a main quality criterion
with glass being the preferential material.
}

\item{Material optical potential $V_F$: 
The coherent neutron scattering length and material density
defines the absolute value of $V_F$ which
determines the energy range 
where total UCN reflection 
under any angle of incidence occurs \cite{Ignatovich1990}. 
A high value of $V_F$ therefore
allows to transmit UCN with higher energies 
and hence increases the UCN intensity.
}

\item{
Neutron losses via material interaction:
UCN reflect from surfaces at any angle of incidence 
in case their kinetic energy is below $V_F$.
As described by an
imaginary part of the potential \cite{Golub1991},
there is a small probability that the UCN undergoes nuclear capture
during reflection
due to the neutron wave-function which slightly penetrates the surface barrier.
The UCN losses are therefore energy dependent.

In addition, the UCN can also inelastically scatter from surface atoms
or impurity atoms sticking to the surface \cite{Atchison2007a}. 
Wall temperatures always exceed UCN temperatures, causing UCN acceleration
out of the UCN regime via phonon scattering.
The overall loss due to these surface effects can be parametrized by a
``loss-per-bounce'' coefficient
as a ratio of the imaginary and real parts of the optical potential. 
This coefficient is independent on kinetic energy. 
From this and the kinetic energy of the UCN one can calculate 
the loss per bounce probability $\mu$ by using Eq. (2.68) in \cite{Golub1991}. 
An energy-averaged loss per bounce probability $\mu_{av}$ 
can be  
estimated by transmission measurements. 
For the extraction of the 
loss-per bounce coefficient 
one needs to know 
the energy spectrum and the angular distribution of the UCN.
}

\item{Gaps: 
The passage of UCN through a guide can be regarded
similar to the propagation of an ideal gas.
Gaps and holes, necessary e.g. for vacuum pumping, 
represent direct loss channels
during UCN transport according to their relative surface area. 
In addition, areas of low $V_F$ which are
directly visible to UCN also
represent leaks, 
e.g.\ at positions where surface coating is missing.
Avoiding gaps and holes is therefore of great importance 
in order to achieve a high UCN transmission.
}

\end{itemize}

\section{The prestorage method to determine UCN transmission}
\label{sec:prestorage}

\begin{figure}
\centering
\includegraphics[width=0.8\textwidth]{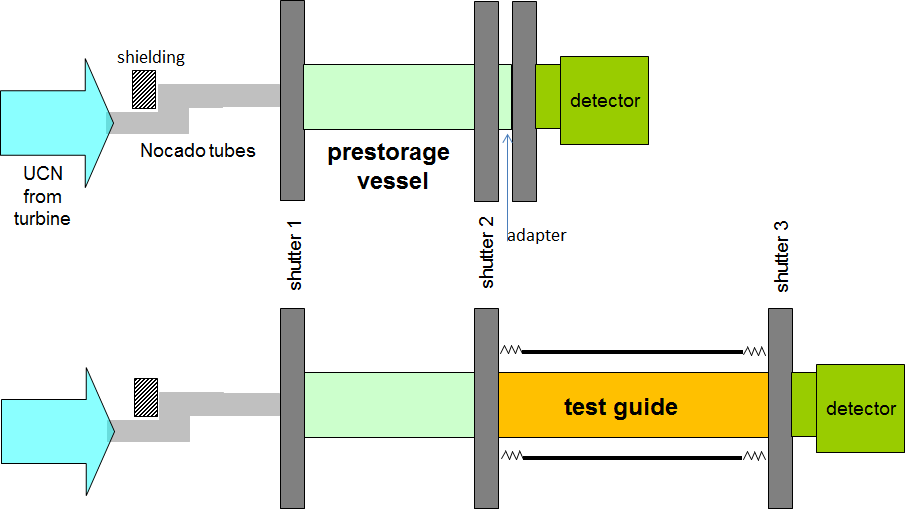}
\caption{
Sketch of the prestorage measurement setup:
a) calibration measurement;
b) test guide measurement.
}
\label{setup}
\end{figure}

The setup for the prestorage measurement 
is sketched in Fig.\ref{setup}.
%
%
A prestorage vessel is filled with 
a defined and known number of UCN in a vessel 
prior to their release into a sample guide or directly into the detector.
These stored UCN can then be directly measured 
with a detector mounted onto the vessel 
and hence be used as calibration.
Then, an additional UCN guide - the one to be tested -
is mounted between the prestorage vessel and the detector 
and the measurement is repeated.
The
comparison of the integrated
UCN counts in the two measurements 
is defined as 
the UCN transmission through the test guide
for the given UCN energy spectrum.

This measurement setup resembles a
small version of the PSI UCN source setup,
where the neutrons may also be
stored in an intermediate storage vessel. 
All UCN with kinetic energies above the material optical potential 
will be rapidly lost during storage, 
defining a reproducible 
energy spectrum.
The prestorage
vessel also influences the momentum direction distribution
due to diffuse reflections. 
UCN passing long UCN guides 
have momenta peaked along the direction of
the guide axis \cite{Ignatovich1990}. 
After a sufficiently long storage period -- of tens of seconds --
this peak is largely reduced.

A similar approach to determine guide transmission
was followed by \cite{Ageron1978} for a longer and
geometrically more complicated UCN guide. 
This measurement and analysis was criticized by \cite{Ignatovich1990} 
as an invalid method. 
The criticism is based on the fact that the prestorage
vessel in the experiment was emptied through a small orifice, thus,
the experiment was mainly comparing the storage time of the vessel,
the outflow time and the measurement time. 
The fact that in
our case the geometry is much simpler, i.e. the diameter of the UCN
guide is not reduced in comparison to the storage vessel and there
are no bends in the setup, makes the method primarily sensitive to the
UCN transmission of the guide. 
Besides a simple analysis,
our measurements can be used
to tune Monte Carlo simulations in order to describe
realistic guide performances \cite{MCUCN2010,Goeltl2012}.

\subsection{Experimental setup}

Our experiment was carried out 
at the 
PF2 facility of the Institut Laue-Langevin (ILL) using the 
EDM beamline of the
UCN-turbine \cite{Steyerl1986}.
UCN are typically guided from the turbine port towards the experiment using electro-polished
stainless steel tubes manufactured by 
Nocado\footnote{Nocado GmbH \& Co. KG, Kirchweg 3, 26629 Groefehn, Germany}. 
The filling line included two 90\degree{} bends, 
which allow for an accurate setup alignment and 
significantly decrease the amount of UCN with kinetic energies above
the material optical potential of steel.
%
The Nocado tubes used have an inner diameter of 66\,mm. 
At the end of the guides, the UCN enter the
prestorage vessel through a stainless steel
adapter flange mounted on customized
vacuum shutters,  
special DIN-200 shutters from 
VAT\footnote{VAT Vakuumventile AG, Seelistrasse, 9469 Haag, Switzerland}
with inside parts coated with diamond-like carbon \cite{Atchison2005c,Atchison2006,Atchison2007a,Heule2008}
(see Sec.\ref{VAT:section}). 
These shutters were later installed as
beam ports of the PSI UCN source.

The measurement setup
consists of a prestorage unit and
a detection unit, 
which both remain unchanged during the measurements.
In the calibration measurement a stainless steel adapter
connects these two units (Fig.\ref{setup}a).
An additional 
test guide is mounted between these units
in a transmission measurement.
(Fig.\ref{setup}b).

The prestorage unit is confined by
shutter~1 and shutter~2.
The storage vessel is a tube made from 
DURAN$^\copyright$\footnote{SCHOTT AG, Hattenbergstr. 10, 55122 Mainz, Germany},
a borosilicate glass, with
180\,mm inside diameter, 5\,mm wall thickness.
The tubes are
sputter-coated\footnote{S-DH GmbH Heidelberg, Sputter-D\"unnschichttechnik,
Hans-Bunte-Strasse 8-10, 69123 Heidelberg, Germany}
on the inside
with about 400\,nm of nickel-molybdenum (NiMo),
at a weight ratio of 85 to 15,
an alloy with a Curie temperature well below  
room temperature \cite{Gosh1998}.
The use of the same surface coating in the prestorage vessel 
as in the guides 
shapes the UCN energy spectrum in a suitable way.

The detector unit consists of a similar VAT shutter (No.~3)
which is only used as a connector unit
to the 2D-200 CASCADE-U detector\footnote{CDT CASCADE Detector Technologies GmbH, 
Hans-Bunte-Strasse 8-10, 69123 Heidelberg, Germany} 
via a 150\,mm long NiMo coated glass guide contained in 
a vacuum housing.
The Cascade-U detector
is a gas electron multiplier (GEM)-based UCN detector using a 200\,nm thin-film 
of $^{10}$B deposited
on the inside of the 
0.1\,mm AlMg3 entrance window 
of the detector to convert UCN to two charged particles 
($\alpha$ and $^7$Li). 
These particles
ionize the detector gas\footnote{Typically we used as
counting gas Ferromix, a mixture of 18\% CO$_2$ and 82\% Ar.}. 
The charge is amplified by the GEM foils 
and detected by a pixelated readout structure. 
The sensitive area of the detector
covers the inner diameter of the glass guides
\cite{Goeltl2012}.

\subsection{Measurement sequence}

Our standard sequence for transmission measurements has the following scheme:\newline
\hspace*{4mm}- Wait for the UCN turbine signal (shutter~1 open, shutter~2 closed);\newline
\hspace*{4mm}- Fill the storage vessel for an optimized filling time of 30\,s;\newline
\hspace*{4mm}- Close shutter~1;\newline
\hspace*{4mm}- Store UCN for a preset storage time of 5\,s;\newline
\hspace*{4mm}- Open shutter~2;\newline
\hspace*{4mm}- Count UCN as a function of arrival time in the detector;\newline
\hspace*{4mm}- Close shutter~2 after the measurement time;\newline
\hspace*{4mm}- Open shutter~1 and wait again for UCN.\newline
Shutter~3 stays permanently open during 
the entire measurement sequence
and functions in transmission measurements only as connector piece.
However, it is used in storage measurements.


\subsection{Shutter properties}
\label{VAT:section}

The large VAT shutters used have opening/closing times of about 1\,s.
The prestorage method is sensitive to the precise timing of shutter operations.
We therefore
measured the timing properties and the shutter
opening function which influences
the path of the UCN and their arrival at the detector.

\begin{figure}
\centering
\includegraphics[width=0.6\textwidth]{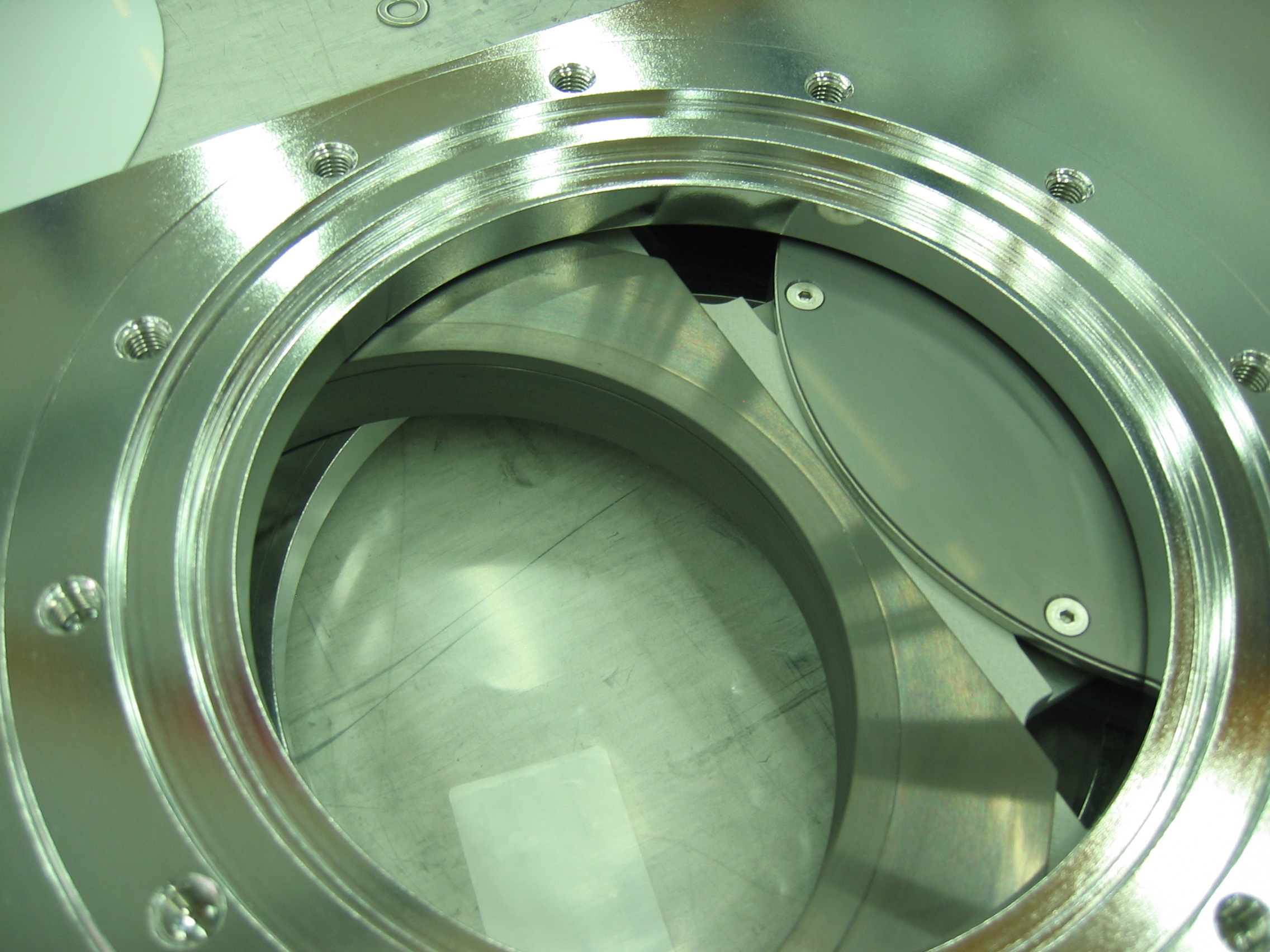}
\caption{
VAT shutter opening 
showing the insert frame 
with opening hole and closing disc
in an intermediate position while moving.
}
\label{vat-image-moving}
\end{figure}

The shutter body is made of aluminum
containing a moving part with a round opening 
and closing disc.
All parts seen by UCN in the open and closed position,
or during movement
are coated with DLC.
When the closing disc is retracted, the 25\,mm gap 
at the center of the shutter is covered by an expanding
ring to close the gap.
Fig.~\ref{vat-image-moving} shows the VAT opening 
displaying the frame of 
the moving part in an intermediate position.
The shutter is air-actuated and controlled with electric valves.

We have measured the timing between the closing of shutter~1 and opening
of shutter~2. 
This time defines the UCN storage time 
and hence the number of neutrons released into the sample.
Its accuracy is crucial for 
the reproducibility of the measurements.

\begin{figure}
\begin{center}
\includegraphics[width=0.75\linewidth]{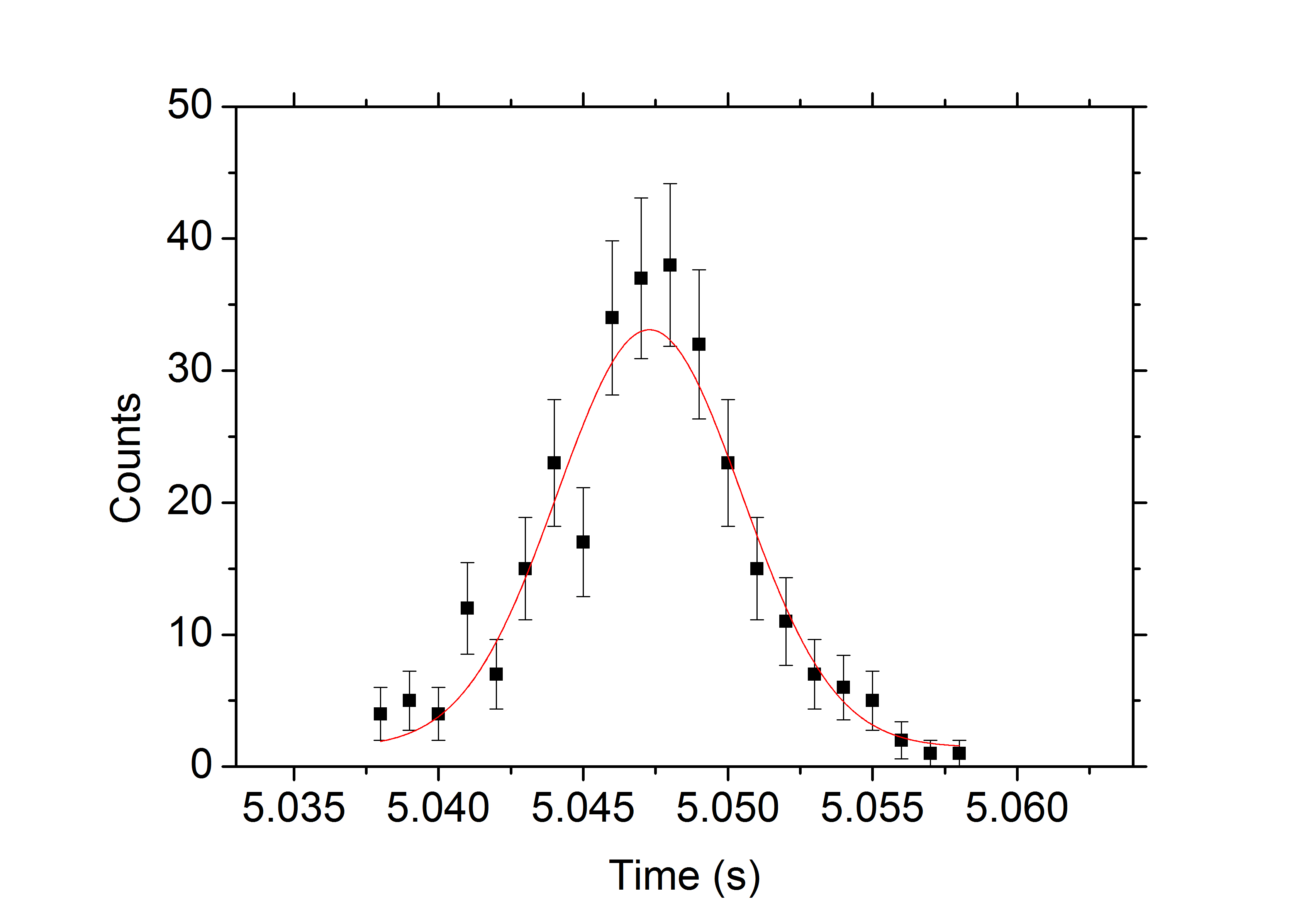}
\caption
{Histogram of the time difference between 
the closed signal from shutter~1 and the open signal from
shutter~2. 
The red line shows a Gaussian fit with a standard deviation of 3.2\,ms.}
\label{VAT:jitter}
\end{center}
\end{figure}

In a separate measurement
with similar conditions concerning 
actuating pressure and environmental temperature 
we determined without neutrons 
the relative timing stability of shutter~1 and 2. 
%
The time 
of the shutter end-switch signals
with respect to the 
slow control start signal was measured.
The resulting time differences were
filled into a histogram with 1\,ms bins,
shown in Fig~\ref{VAT:jitter}. 
Note that the experiment was set to have 
exactly 5\,s time difference which was accurate on the 10$^{-3}$ level.
The standard deviation $\sigma$ 
of a Gaussian fit is 3.2\,ms, reflecting an
excellent reproducibility of the opening and closing times.

\begin{figure}
\begin{center}
\includegraphics[width=0.6\textwidth]{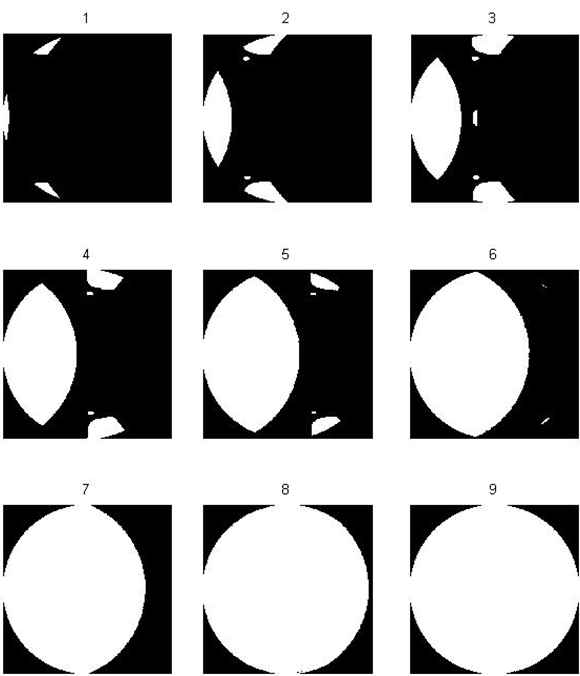}
\caption{
Nine selected pictures showing the opening of the VAT shutter.
Opening function of the VAT shutter, 
1 - just after opening;
3 - the main obstructive part of the shutter holder 
is visible in the center of the opening;
9 - the shutter is fully open.
}
\label{VAT-opening-frames}
\end{center}
\end{figure}

\begin{figure}
\begin{center}
\includegraphics[width=0.7\textwidth]{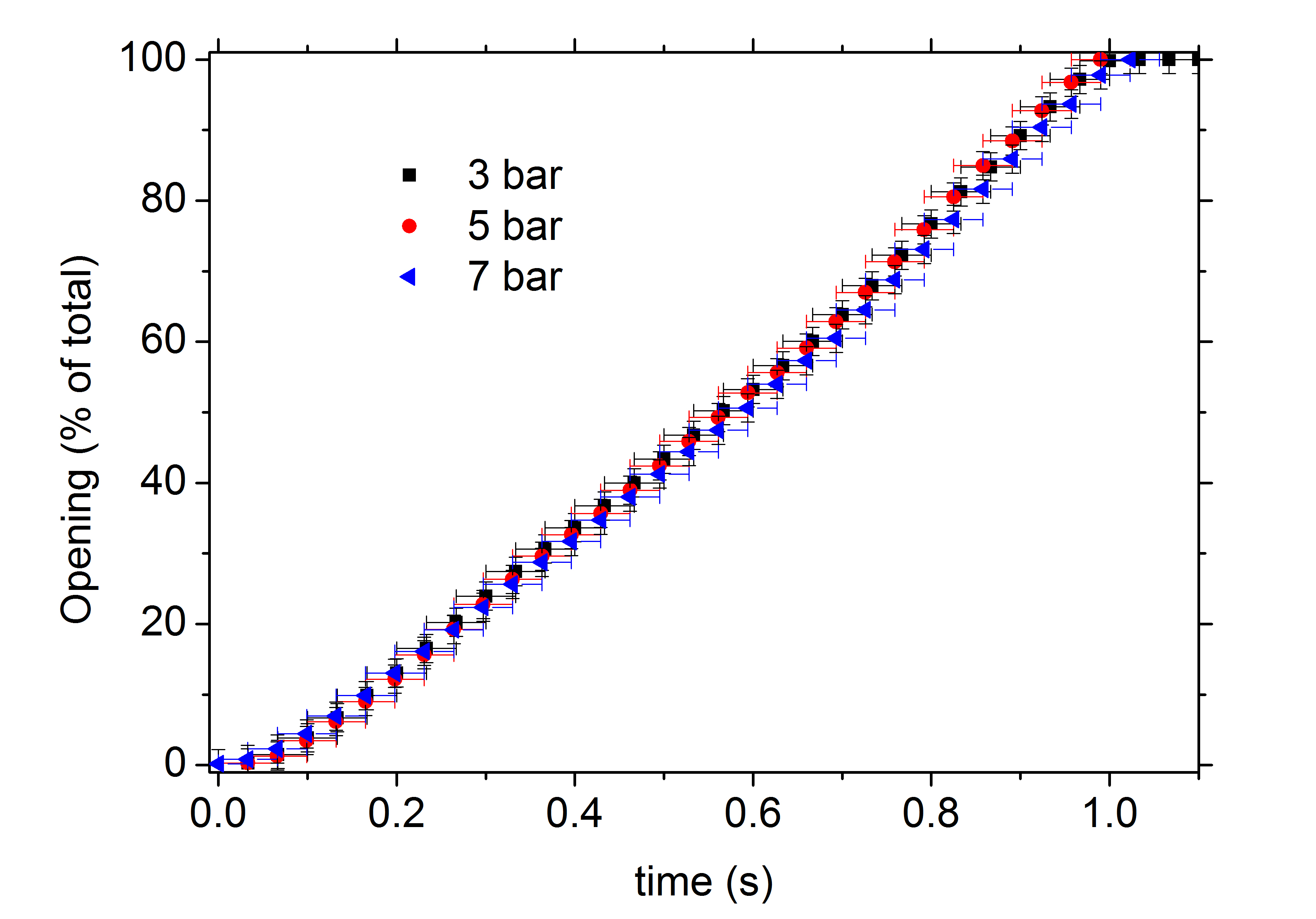}
\caption{
Opening function of the VAT shutter 
-- given in percent of total opening --
measured with three different actuator pressures
of 3, 5 and 7\,bar.
}
\label{VAT-opening}
\end{center}
\end{figure}

The opening of the shutter also partially obstructs the path of the UCN
during the movement, which is reflected in the opening function.
We used a bright lamp and a camera for the measurement of the light transmission 
passing or deflecting on the intercepting shutter parts,
supposing a comparable opening function for light and UCN.
Fig.\ref{VAT-opening-frames}
shows a selected sequence of pictures 
showing the 
shutter during opening.
Fig.\ref{VAT-opening}
shows the resulting opening function measured 
with three different actuator pressure settings of 3, 5, and 7\,bar.
No significant difference was observed for the different pressures.

\clearpage

\subsection{Calibration setup}

The standard calibration setup is used to determine the
UCN transmission through the setup
without a test guide.
The detection unit is connected to the prestorage
vessel using a special adapter,
a 60\,mm short stainless steel piece.
In order to optimize the
transmission properties for the UCN and to minimize the influence of
this adapter on the measurement, it was made as short as
possible and the inner surface was hand-polished. 

In addition, the standard setup
was modified in such a way that shutter 3 and
the stainless steel adapter were removed. 
By comparing the results
from the standard calibration measurement and the modified
calibration measurement one obtains the influence of the stainless
steel flange on the total count rate. 
In section~\ref{calibration:section} 
we show these measurements
agree within statistical uncertainties. 
Hence, the influence of the adapter can safely be neglected.\newline

\subsection{Setup for transmission measurement}

\begin{figure}
\begin{center}
\includegraphics[width=0.85\textwidth] {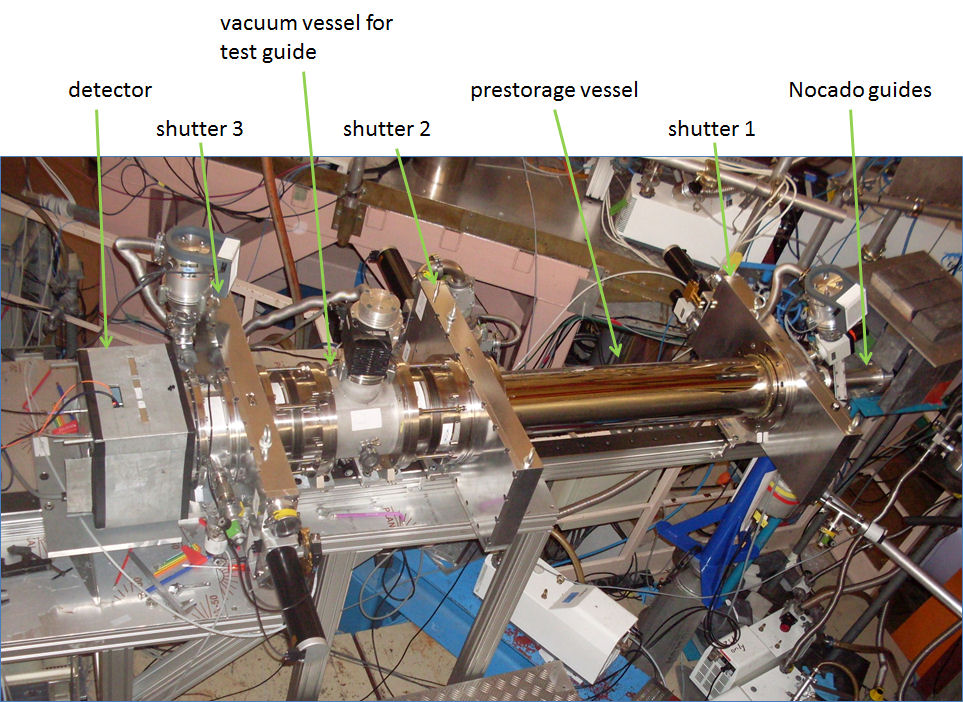}
\caption
{Photograph showing the setup installed at the EDM beamline at ILL
for the measurement of the guide 1S3. 
The Cascade detector (in cadmium shielding) 
with shutter-3, 
the prestorage vessel with shutter 1 and 2. 
Behind shutter~1 are the Nocado filling guides hardly visible.
}
\label{ILL-setup-foto}
\end{center}
\end{figure}

A photo of the setup installed at the ILL EDM beamline 
for the transmission measurements
is shown in Fig.\ref{ILL-setup-foto}.
The test guides were mounted in a custom vacuum housing between
shutters~2 and 3.
As all the guides were designed to be installed at the PSI UCN source 
short adapter pieces had to be manufactured in order to connect the guides 
to the VAT shutters in the transmission measurements. 
All adapter pieces were made of stainless steel
with a maximal length of 40\,mm.
The inside surfaces
which act as neutron guides were hand polished 
to have 
negligible influence on the transmission measurements.

Care was taken to minimize any possible gaps between guides and adapters.
Two flexible bellows at 
both ends of the vacuum housing setup allowed to adapt
the vacuum housing length 
to the total guide and adapter length.

\begin{figure}
\begin{center}
\includegraphics[width=1.1\linewidth]{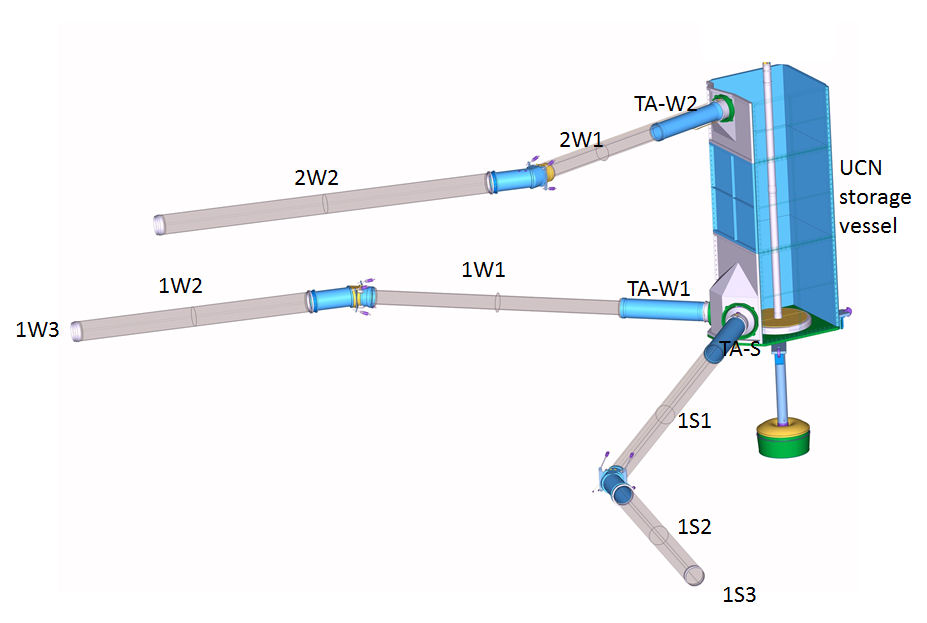}
\caption{Drawing of some parts of the UCN source relevant for
this paper, 
showing the UCN storage vessel (about 2.5\,m high)
and the three UCN guide sections towards the experimental ports
West-1, West-2 and South with the 
naming scheme for the corresponding guide parts.
The green vessel displays the container for
the solid deuterium used for UCN production.
All the shown parts are contained in a large vacuum tank \cite{Lauss2012,Lauss2014}
}
\label{fig:guide-names}
\end{center}
\end{figure}

Table~\ref{guide:lengths} states names, lengths and materials of 
the measured guides.	
Their names refer to the subsequent placement 
at the PSI UCN source which is shown in Fig.\ref{fig:guide-names}.
The given guide lengths include the stainless steel end flanges 
which are permanently 
glued to the glass guide \cite{Bertsch2009} to allow for a
stable connection and minimal gap widths between
various parts 
in the final installation at PSI.
In addition, the total length includes the additional stainless steel adapter.

\begin{figure}
\begin{center}
\includegraphics[width=0.8\linewidth]{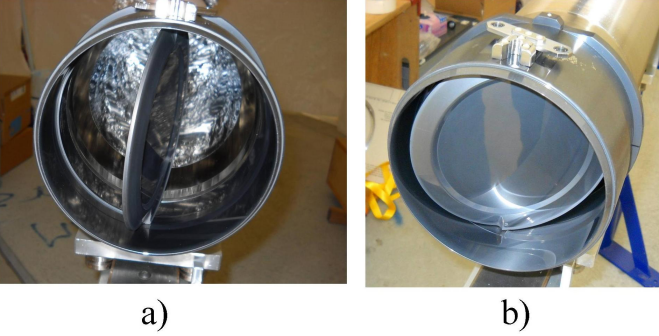}
\caption{
View of the UCN butterfly valve in open (a) and 
half closed (b) position.
} 
\label{butterfly-valve}
\end{center}
\end{figure}

All guides have inner diameters of 180\,mm,
only the guides 2W1 and TA-W2 
have inner diameters of 160\,mm.
All guides, glass and stainless steel, were coated on the inside with the
same NiMo coating with a weight ratio of 85 to 15 percent which is non-magnetic at room temperature.
The small section with the UCN butterfly valve shown in Fig.~\ref{butterfly-valve}
is coated 
with diamond-like carbon (DLC).
Guides 1S1 and 1W3 are similar guides with identical dimensions and properties.

\begin{table}
\begin{center}
\begin{tabular}{|c|c|c|c|c|c|}
  \hline
  Guide name & Material & Total length & Guide length & Glass length & Inside diameter  \\  \hline
   &  &  (\,mm) & (\,mm) & (\,mm)  & (\,mm)  \\  \hline \hline
	1W1 	& glass	& 2624 		& 2565.7 		& 2498.7	& 180	\\	\hline
	1W2		& glass	& 2351 		& 2344 			& 2320	& 180	\\	\hline
	2W1 	& glass	& 1655 		& 1597 			& 1530	& 160	\\	\hline
	1S1 	& glass	& 3783 		& 3725 			& 3658	& 180	\\	\hline
	1S2 	& glass	& 2628 		& 2621 			& 2597	& 180	\\  \hline
	1S3   & glass 	& 603  		& 575 	  & 543		& 180 \\	\hline \hline
	TA-W1 & st.steel  & 1024  & 900    &  - &  180  \\ \hline
	TA-W2 & st.steel  & 1024  & 900    &  - &  160  \\ 
	\hline
\end{tabular}
\end{center}
\caption{
	Guide names and corresponding lengths. Names refer to locations
	for final mounting in the UCN source setup. 
	The given guide lengths
  include the NiMo coated stainless steel flanges glued onto the glass guides. 
	The total lengths include also the stainless steel adapter pieces 
	necessary to mount the guides in the setup. 
	The material column defines the material of the tube, namely glass or
	stainless steel. The stainless steel guides include the part with the 
	neutron valve.
	1S1 and 1W3 are similar guides with identical dimensions and properties.
}
\label{guide:lengths}
\end{table}

Due to the large dimensions of the setup and the high weight of the
components, accurate alignment was critical in order not to
damage the equipment.
The prestorage unit,
all vacuum vessels for the tested guides,
the VAT shutters, 
and the detector unit were all mounted on 
custom made carriages 
that could be moved on a 
Hepco GV3 rail system~\footnote{Hepco Motion
Lower Moor Business Park, Tiverton, 
Devon, United Kingdom}.
Thus, a precise
alignment of the components with respect to each other
could be achieved.

All measurements were performed at vacuum pressures  
below 10$^{-3}$\,mbar.


\section{Optimization of the measurement}
\label{optimization} 

To investigate and optimize the experimental conditions
systematically it is
useful to divide it into three steps which are repeated 
several times in a cycle:
\begin{enumerate}
\item{Filling UCN into the prestorage vessel.}
\item{Storing UCN in the prestorage vessel.}
\item{Releasing the UCN out of the vessel via the test guide to the detector.}
\end{enumerate}
In order to start every measurement from the same initial
conditions shutters~1 and~2 were closed before step~1.
Triggered by the turbine signal the opening of shutter~1 
started the filling phase, thus allowing the UCN density in the 
storage vessel to build up. 
Shutter~1 closed after the filling time.
During the subsequent storage phase
the UCN energy spectrum shifts
to a lower mean value due to velocity dependent
losses, i.e.\ for 
UCN energies larger than the material optical potential of the material
trap and 
because of losses on the material surface 
(loss-per-bounce) 
at different bounce rates
\cite{Golub1991,Ignatovich1990}.

Once the storage time had elapsed, shutter~2 was opened,
allowing the remaining UCN to move freely in the combined volume of
the prestorage vessel, the test guide, and the detector unit
where they are counted. 
Although, most UCN penetrate the AlMg3 detector window, 
a small fraction is initially
reflected
and may be counted at later times.

\subsection{Filling time}
\label{filling:time}

In order to maximize the amount of UCN per cycle
the filling time of the prestorage vessel was optimized.
We define the filling time between the beginning
of the filling process, i.e.\ the signal from the turbine, and the
electronic signal to the shutter triggering its closing motion.
After a scan of different shutter~1 closing times
30\,s was
chosen as the working point for all measurements.

In order to understand possible systematic effects in the filling phase, 
one has to pay attention to the fact 
that the UCN turbine at ILL is a multi-user facility \cite{Steyerl1986}. 
It serves three main beam ports sequentially, i.e.\ there is a UCN
guide inside the turbine which is moved by a stepping motor towards
the port that has requested UCN. 
Once the guide is pointed at one
port, the user working at that port obtains an electronic signal to
confirm the guide position. 
However, the UCN density in the guide-part up to the port
builds up with time, thus, making the UCN density up to
shutter~1 a function of the position of the distributing guide prior
to the electronic signal.

In the process of the data analysis we noticed a difference in the
number of UCN depending on the position of the turbine prior to the
beginning of a measurement cycle. The effect was noticed twice but
may have happened more often. 
For the known occasions where the
turbine pattern changed during a measurement period the effect
amounts to 0.007 of the total number of UCN detected. 
The data
acquisition system used for the experiments added the data of
subsequent cycles until stopped by the user, alas, most measurements
were done in long over-night runs without attendance. 
Thus, the data
may contain cycles with either reduced or increased numbers of UCN.
Due to the fact that all the cycles are superimposed
of the DAQ system used in these runs, it is
impossible to separate the cycles and the turbine position imposes a
relative 
systematic uncertainty of up to 0.007 on all measurements. 
This uncertainty
dominates the uncertainty budget of the entire experiment. 
As will be shown later in section~\ref{Results:Section} 
the transmission values
are consistent and the uncertainty described here could well be an
overestimation.

\subsection{Storage time}
\label{storage:sec}

The duration of the storage phase determines the
shaping of the energy spectrum of the UCN that will be measured.

\begin{figure}
\begin{center}
\includegraphics[width=0.95\linewidth]{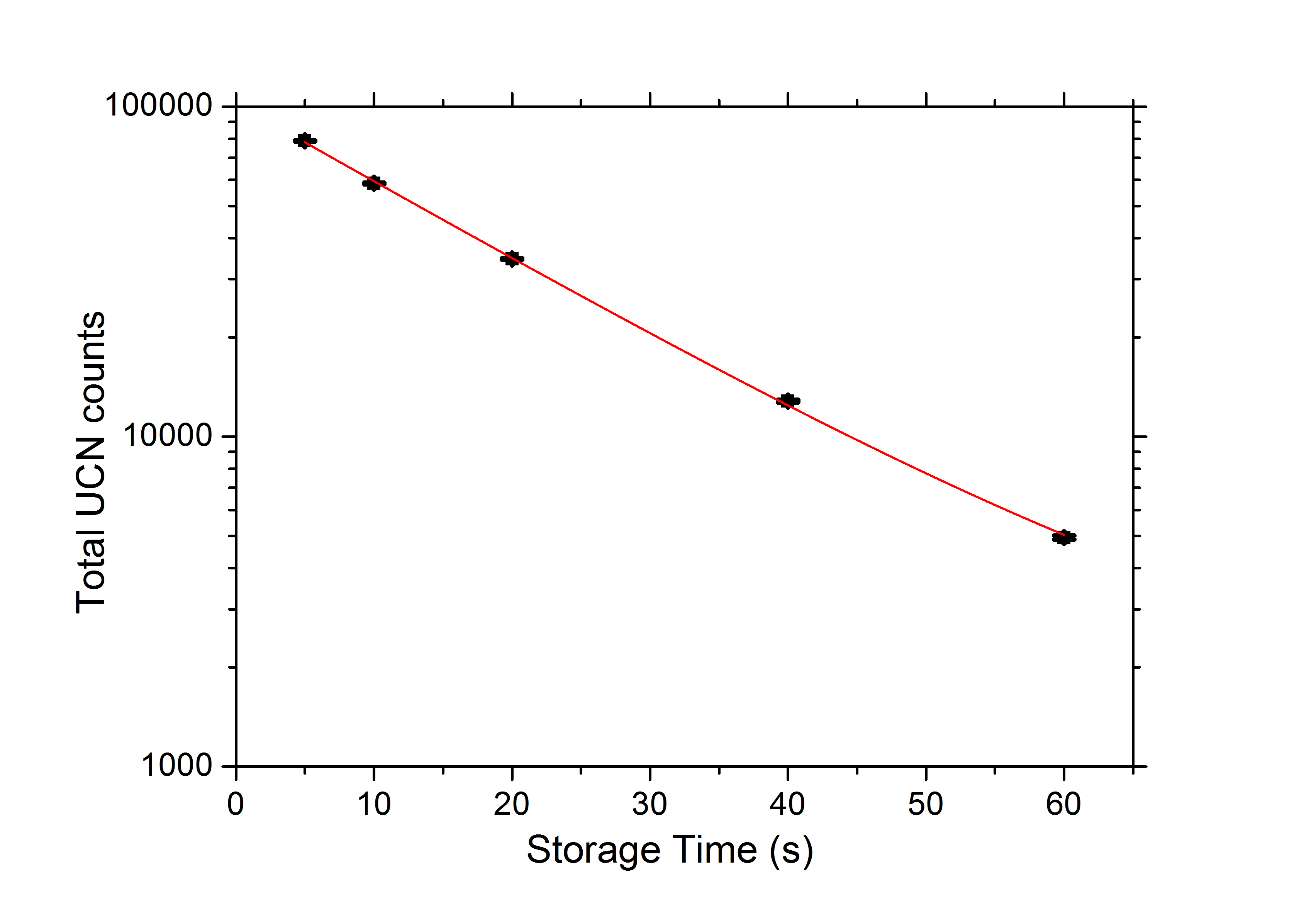}
\caption{
The total UCN counts as a function of the
storage time in the prestorage vessel. 
A fit with a single exponential
results
in a storage time of 18.5$\pm$0.5\,s.
} 
\label{storagetime:prevolume}
\end{center}
\end{figure}

To quantify the storage properties of the prestorage vessel
the storage time constant (STC) was determined
from a single exponential fit to the data.
We define the
storage time to be the time that elapses between the electronic
signal closing shutter~1 and the signal opening shutter~2. 
Figure~\ref{storagetime:prevolume} shows the measurements together 
with a single exponential fit
resulting in a STC of 18.5$\pm$0.5\,s.
The storage curve is rather poorly defined by 
the fit because 
it neglects the energy dependence of UCN storage.
%
However, for an estimation of the
systematic uncertainty it is sufficient.

In order to estimate the systematic
uncertainty of the number of stored UCN 
we use the STC from the single exponential fit
and the measured rate of the filling time
which is 
$\sim$130\,UCN/s at 30\,s of filling time.
The relative uncertainty of 0.053\,s$^{-1}$
from the measured filling rate
is a function of the timing uncertainty 
mainly due to 
control electronics and shutter movement.
Using FWHM = 7.6\,ms of the shutter timing
as uncertainty of the storage time 
the corresponding relative uncertainty in UCN counts
computes to 
0.053 $\times$ 0.0076 = 4$\cdot$10$^{-4}$.

\subsection{Emptying the prestorage vessel}

\subsubsection{The calibration measurement}
\label{calibration:section}

\begin{figure}
\begin{center}
\includegraphics[width=0.6\linewidth]{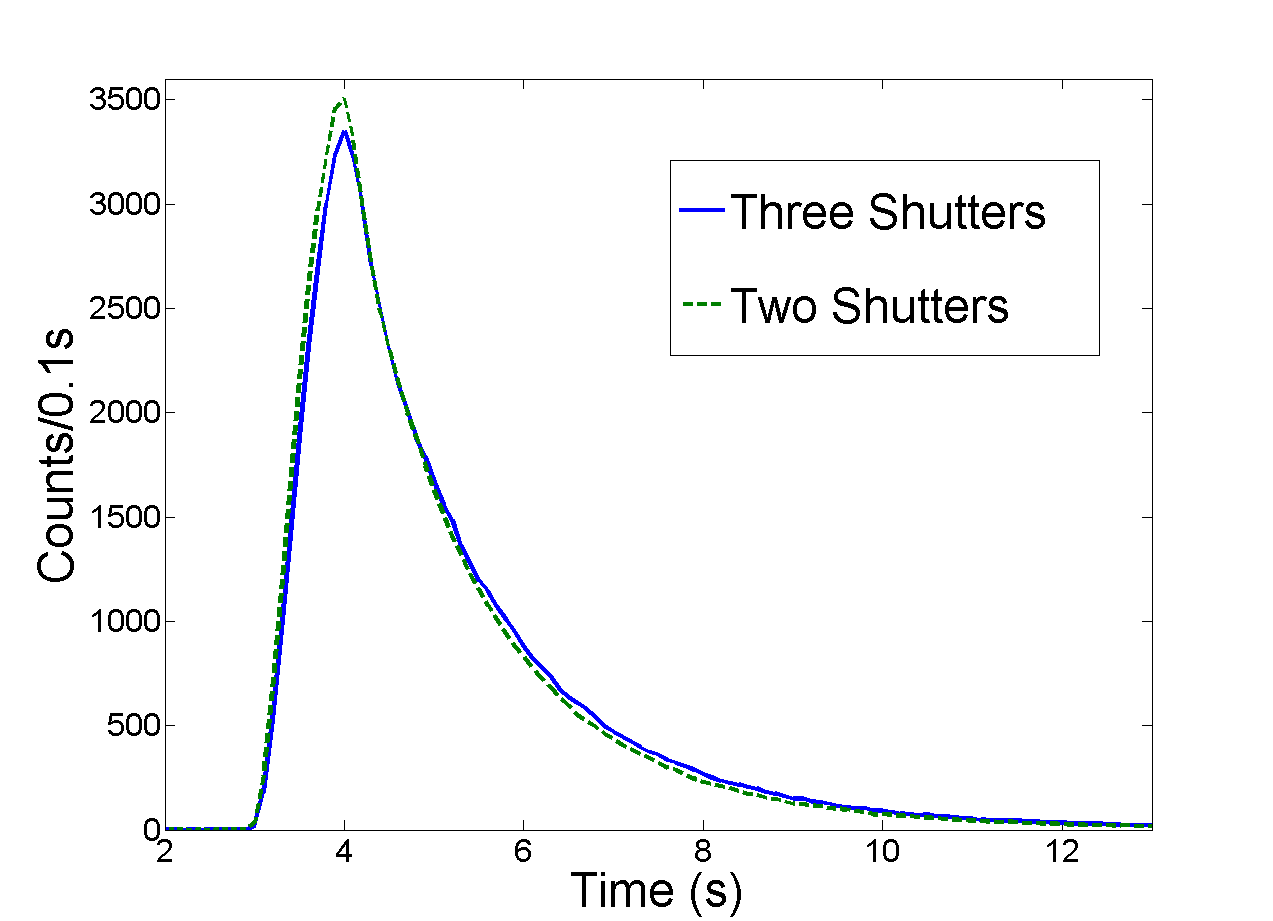}
\caption{
Comparison of emptying curves for the two 
setups,
calibration setup with 3 shutters and adapter 
(standard calibration setup),
and calibration setup with 
shutter~3 removed.
The plotted UCN counts after shutter opening 
shows the sensitivity of the setup
towards a rather small change in length
which causes a small change in arrival time.
} 
\label{comparison:calibs}
\end{center}
\end{figure}

To investigate the influence of the stainless steel adapter used in
the calibration setup a dedicated measurement was performed.
The detector unit was 
mounted directly onto shutter~2,
removing 
shutter~3 and the stainless steel adapter.
However, this largely increases the time necessary for a setup change.
Figure \ref{comparison:calibs} shows 
a comparison of the 
time distribution of UCN counts 
arriving 
after shutter~2 opening.
The slight difference in the shape of the two curves can be explained by
a 30$\%$ difference in length of two flight paths
behind shutter~2. 
Using all recorded measurement cycles 
the mean values 
of the total number of UCN detected in the modified setup,
69595$\pm$52, 
and in the standard calibration setup,
69577$\pm$60, 
agree within statistical uncertainty.
One can conclude that the influence of the
stainless steel adapter on the obtained transmission values is
negligible.

\begin{figure}
\begin{center}
\includegraphics[width=0.60\linewidth]{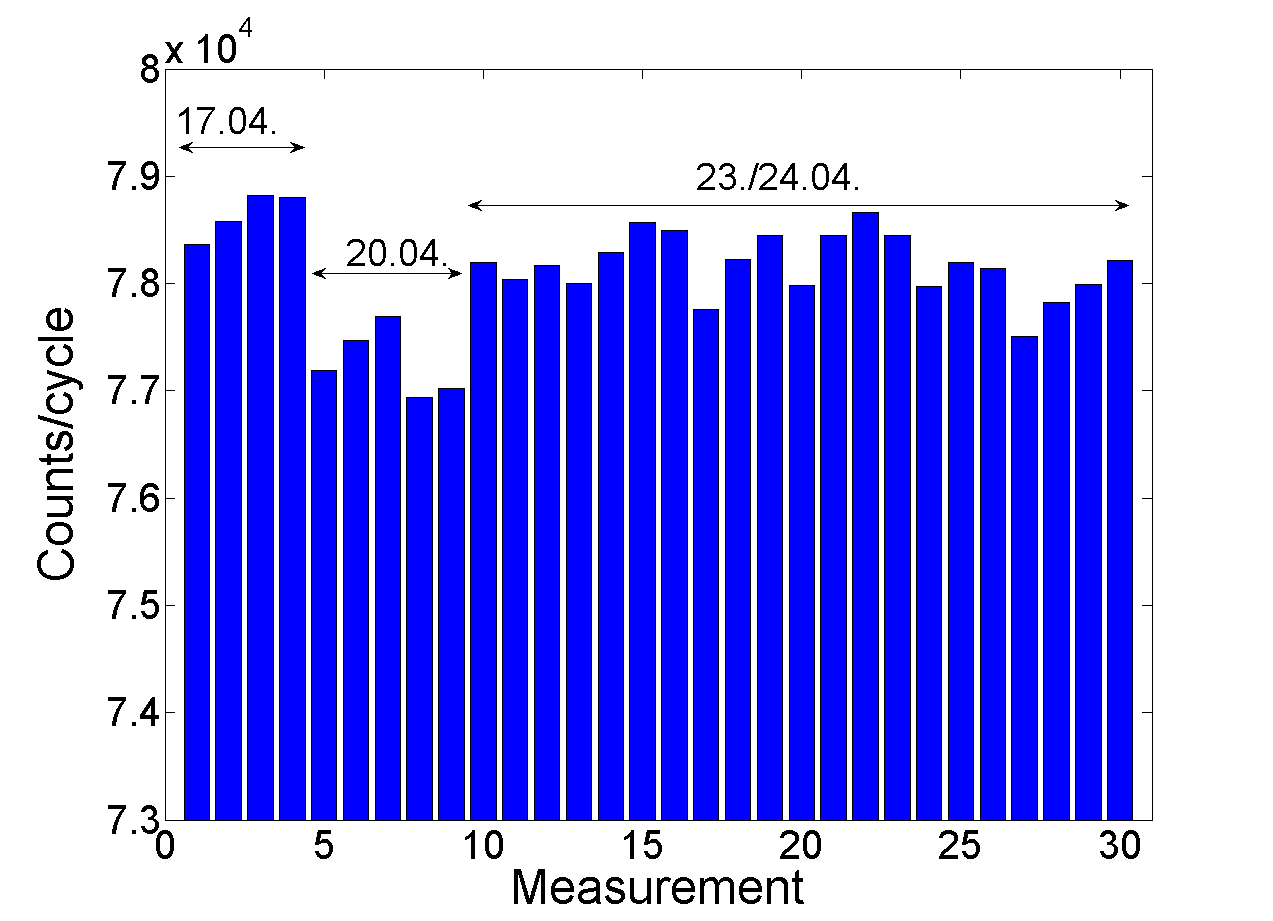}
\caption{
Mean number of UCN over time for different calibration measurement cycles.
The data was taken on
four different days, 
in between the indicated dates the experiment
was disassembled and reassembled, thus indicating the
reproducibility of the obtained results. 
Measurements from
20.04. were taken under inferior vacuum conditions 
and not further used in the analysis.
} 
\label{prestorage:calibs}
\end{center}
\end{figure}

In order to check for systematics and
long term drifts the calibration setup was (re)assembled and
measured at several occasions throughout the campaign. 
The results
from the measurements performed on the 17.04. and
23.-24.04., as shown in Fig.~\ref{prestorage:calibs}, 
differ by 0.4$\%$ 
in the mean number of UCN counted per cycle.
Fluctuations of UCN intensity due to changes in cold source
performance or reactor power of 0.5$\%$ are not unusual.
In between the
two measurements the entire setup, except the detection unit, 
was taken apart and reassembled. 
The agreement of the two
measurement periods demonstrates the level of reproducibility of
the setup. 
The measurements taken on the 20.04. were taken under
vacuum conditions in the 10$^{-2}$\,mbar range. 
This data points towards the necessary vacuum conditions for measurements
and
is not used for the calculation of the mean number of UCN per
calibration cycle.

Averaging over all calibration measurements performed on April 17, 23,
and 24, one obtains a mean number of 78292$\pm$68 UCN
per calibration cycle
used in the further analysis.

\subsubsection{Transmission measurements of glass guides}
\label{glass-guides}

All measurements were taken with the Cascade-U detector in
time-of-flight (TOF) mode, i.e.\ the slow-control started the detector one
second before shutter~2 was opened. 
The emptying curve of the
experiment was recorded with a time resolution of 1\,ms
and re-binned by a factor of 10 in the analysis.

\begin{figure}
\begin{center}
\includegraphics[width=1.10\linewidth]{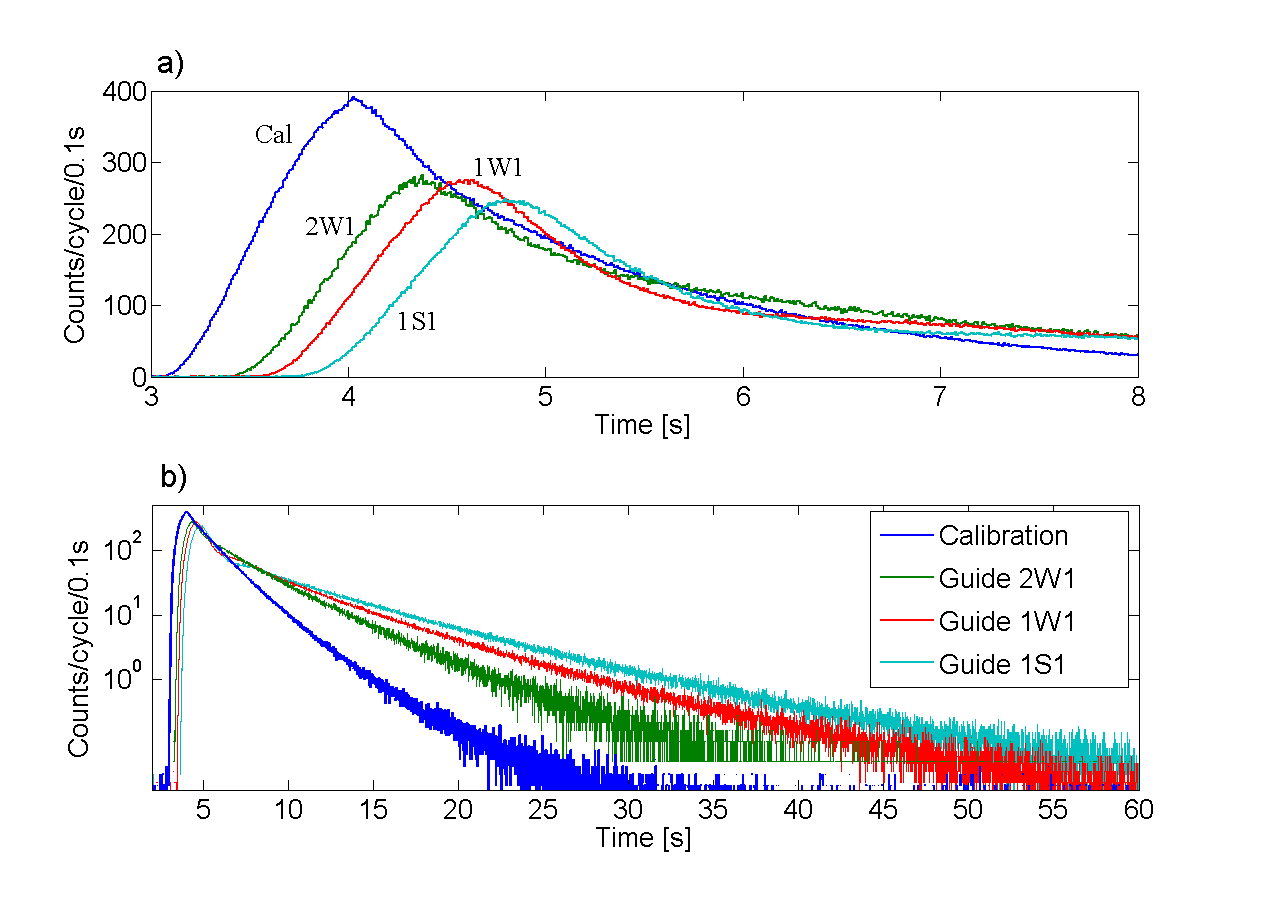}
\caption{
Glass guide measurements:
Emptying time spectra
for the calibration setup and the 
transmission setup with guides
1S1, 1W1 and 2W1.
a) up to 7\,s after 
the beam kick;
b) up to 60\,s.
The spectra
show a prompt peak which contains most of the UCN and a long tail
which contains UCN which have undergone many reflections. 
In the longest guide UCN are stored up to one
minute before being counted in the detector.
}
\label{alllengths:plot}
\end{center}
\end{figure}

Figure \ref{alllengths:plot} shows emptying curves for different glass guide lengths.
One can clearly see the effect of
increasing distance between prestorage vessel and 
detection unit: 
the rising edge of the spectrum starts at a later
point in time and becomes less steep; the falling edge becomes
longer with the length of the guide.

\subsection{Transmission measurements of stainless steel tubes}

\begin{figure}
\begin{center}
\includegraphics[width=0.6\linewidth]{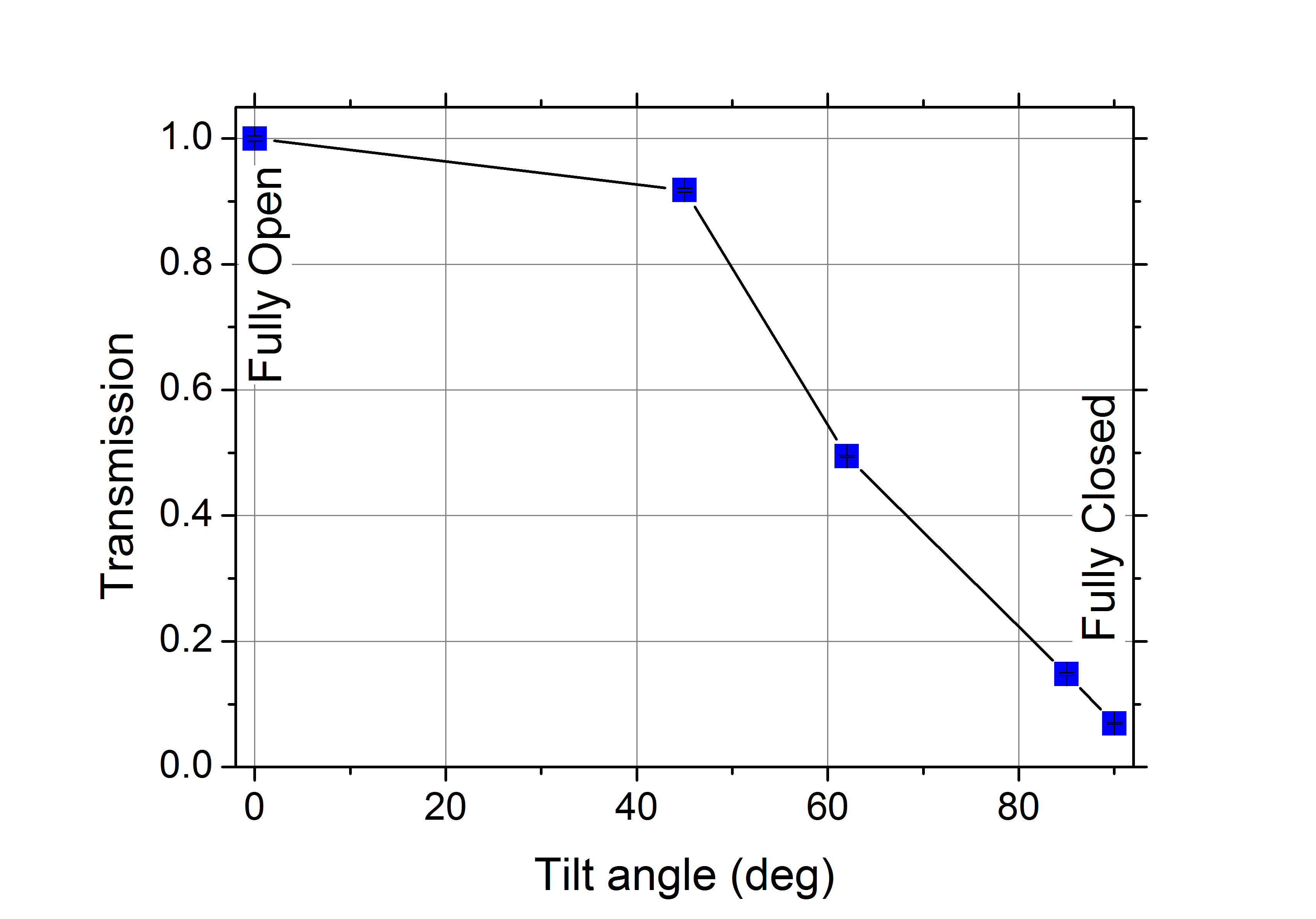}
\caption
{
The dependence of UCN counts on the tilt angle of the UCN butterfly valve 
The valve is rotated around its axis in
the middle of the guide, hence, when fully open only the thickness
of the disc ($\sim$3\,mm) blocks the path of the neutrons over the
entire guide diameter.
Even when fully closed a few percent of UCN leak through the 
not perfectly closing valve. 
Statistical errors are smaller than the symbol size.
The line is for eye-guiding only.
} 
\label{open:closedvalve}
\end{center}
\end{figure}

The first meter of each of the three UCN guides at the PSI UCN source, 
dubbed ``TA'', 
is made of hand-polished and NiMo coated stainless steel \cite{Lauss2012,Goeltl2012}.
It also contains a DLC-coated UCN butterfly valve
shown in Fig.\ref{butterfly-valve}
which can control the neutron flux
towards the experimental area.
Two of the three stainless steel units (West~1 and West~2) 
were tested at ILL in the same way as the guides. 
The measurements were done twice
for each unit with varying tilt angle of the UCN valve. 
Figure \ref{open:closedvalve} shows the dependence of the UCN counts on the
angle of the UCN butterfly valve.

It is interesting to see that up to a tilt angle of $\sim$45\degree,
the decrease of the UCN counts only amounts to $\sim$10\%, while
the ``covered'' cross-sectional area at this angle
corresponds to about 70\% of the entire guide.

\subsection{Normalized UCN transmission}
\label{transmission:subsection}

To allow for a quality check of the guide's surfaces and
coatings and a better comparison between the individual guides, 
the transmission
per meter of guide $T_{\text{PM}}$ is calculated as follows:
\begin{equation}
T_{PM} = \left(\frac{Counts_{\text {Guide}}}{Counts_{\text{Calibration}}}\right)^{\frac{1}{L}}
\label{Transper:meter}
\end{equation}
where $Counts_{\text{Guide}}$ and $Counts_{\text{Calibration}}$ are
the mean total number of counts detected per measurement cycle with
the test guide or the calibration setup respectively, and $L$ is the
length of the tested guide in meters.


\section{Results and Discussion}
\label{Results:Section}

\subsection{Results from the individual measurements}

All measurements were taken in the
time-of-flight mode, however, the transmission is calculated from
the mean total of UCN counts in a given setup, i.e.\ the integral of
the TOF spectrum. 
Table \ref{results:table} gives the average
numbers of UCN counted per cycle for the respective setups. 
The uncertainty given in Tab.~\ref{results:table} is the statistical
uncertainty. 
The last column states the number of cycles used for
the calculation of the mean and standard deviation.
The larger cycle numbers occurred in undisturbed overnight runs.

\begin{table}
\begin{center}
\begin{tabular}{|c|c|c|}
  \hline
  Setup name &  Counts per cycle  & number of cycles\\\hline
  Calib. April & 78292$\pm$68 &  19  \\\hline
  1W1    & 75791$\pm$91    &  8  \\\hline
  1W2    & 76719$\pm$149   &  3  \\\hline
  2W1    & 75763$\pm$68    &  18  \\\hline
  1S1    & 74098$\pm$216   &  4  \\\hline
  1S3    & 76323$\pm$80    &  3  \\\hline
  Calib. Oct.$^{*)}$ & 74927$\pm$48 &  42 \\\hline
  1S2$^{*)}$ & 74817$\pm$38 &  60  \\
  \hline
\end{tabular}
\end{center}
  \caption	
	{Measurement results for glass guides: 
	Mean number of counts per cycle 
with statistical uncertainty only,
	obtained with the different setups.
	The measurements marked with $^{*)}$ were performed in October.
	The difference in calibration counts 
	points towards a different performance of the UCN turbine or feeding line, as the reactor power
	was within 0.5$\%$ the same as in April.}
  \label{results:table}
\end{table}

Table~\ref{stainless:results} shows the results from the stainless
steel guide measurements.

\begin{table}
\begin{center}
\begin{tabular}{|c|c|c|}
  \hline
  Setup name &  Counts per cycle    & number of cycles\\\hline
  Calib. September    & 70613$\pm$24   &  138 \\\hline
  TA-W1 open   				& 69592$\pm$37   &  62  \\\hline
  TA-W1 closed 				& 3888$\pm$8    &  32   \\\hline
  TA-W2 open   				& 68678$\pm$22   &  182  \\\hline
  TA-W2 closed 				& 4779$\pm$21   &  14   \\
  \hline
\end{tabular}
\end{center}
  \caption
	{Results for stainless steel guides: 
  Mean number of UCN counts for the given setups.
  The difference in calibration counts 
	points towards a different performance of the UCN turbine or feeding line, as the reactor power
	was within 0.5$\%$ the same as in April.
}
  \label{stainless:results}
\end{table}

\subsection{Estimate of the systematic uncertainties}

There are several sources for systematic uncertainties in this
experiment.

\begin{itemize}

\item{
The timing uncertainty can be estimated conservatively 
from our measured filling rate of $\sim$130\,UCN/s at 30\,s of filling time.
Hence, even the large assumption of 
a UCN shutter timing jitter of 1\,s 
results in a relative uncertainty of the total number
of UCN per cycle of $<$0.0002.

The DAQ and electronics for the shutter movement
have been tested separately and have shown to be reproducible to a
level of better than 10\,ms 
(Fig.\,\ref{VAT:jitter})
resulting in a relative systematic
uncertainty on the level of 10$^{-5}$.}

\item{
As described in section~\ref{filling:time} the effect of the turbine
UCN guide moving to different positions imposes a relative uncertainty of
0.007 on all measurements. 
However, this is an upper limit as the 
measurements typically average over several different positions of
the turbine UCN guide.
}

\item{
The relative systematic uncertainty arising
from the duration of the
UCN storage amounts to
4$\cdot$10$^{-4}$ (Sec.~\ref{storage:sec}). 
}

\item{
We estimate the relative systematic uncertainty 
from the reproducibility of the calibration measurement 
from Fig.~\ref{prestorage:calibs}
to be on the order of 0.004. 
}

\item{
Due to the fact that the glass tubes used
are not
machined to very high precision, gaps
with typical sizes of 0.1 to 0.2\,mm,
occurred in the guide measurement setup. 
Accurately measuring the impact of those gaps 
on the UCN counts is very difficult. 
In Sec.~\ref{calibration:section} it was demonstrated that a 
60\,mm stainless steel adapter and a complete VAT shutter can be
removed without affecting the number of detected UCN, 
hence, we concluded that estimated gaps 
outside the storage vessel
of about 0.5\,mm,
over the entire circumference of the guide, 
have a negligible influence on our transmission result.
}

\item{
The reactor power is monitored and
was stable to $\sim$0.2$\%$ 
over the measurement period.
Hence, the influence on the measurement results is negligible.
}

\end{itemize}

After carefully analyzing the data, we quote the results with
separate statistical and systematic uncertainties. 
The systematic uncertainty is dominated by the turbine positioning. 
All relative contributions summarized in table \ref{systematics:table}
were added quadratically (0.0081) 
and then increased to 0.0085 for a conservative estimate. 
A relative systematic uncertainty of 0.012 follows for
all quoted transmission values.

\begin{table}
\begin{center}
\begin{tabular}{|c|c|}
  \hline
  Source & Impact \\\hline
  Filling stop time & $<10^{-5}$\\\hline
  Turbine position & 0.0070 \\\hline
  Storage phase & 0.0034 \\\hline
  Calibration & 0.0040 \\\hline\hline
  Total & 0.0081 \\\hline
\end{tabular}
\end{center}
  \caption{
	Summary of estimated values for considered relativ systematic uncertainties
	on the measured count integrals.
	}
  \label{systematics:table}
\end{table}

\subsection{UCN transmission using the calibration measurement}

Using the numbers from Tab.~\ref{results:table} and Tab.~\ref{stainless:results} 
we obtain the transmission values of Tab.~\ref{transmission:values}
for the
UCN spectrum emerging from the prestorage vessel after 5\,s of storage.

\begin{table}
\begin{center}
\begin{tabular}{|c|c|c|}
  \hline
  Guide name  & Total transmission & Transmission per meter \\\hline
  1W1    &  0.968(1)(12) & 0.988(1)(12) \\\hline
  1W2    &  0.980(2)(12) & 0.991(2)(12) \\\hline
  2W1    &  0.968(1)(12) & 0.980(1)(12) \\\hline
  1S1    &  0.946(3)(12) & 0.986(3)(12) \\\hline
  1S2    &  0.999(1)(12) & 0.999(1)(12) \\\hline
  1S3    &  0.975(1)(12) & 0.959(1)(12) \\
  \hline
\end{tabular}
\end{center}
  \caption{Transmission values for the glass UCN guides. 
The uncertainties given are statistical and systematical respectively.}
  \label{transmission:values}
\end{table}

The results show a remarkable performance of the glass guides.
Assuming the guides to have equal properties except their length one
can calculate a mean transmission per meter of 0.989$\pm$0.012.

The transmission values obtained for the stainless steel guides are
given in Tab.~\ref{stainless:trans}.
In case of TA S-1 only the straight guide without the valve was measured
and behaved as TA W-1 given in Tab.~\ref{stainless:trans}.

\begin{table}
\begin{center}
\begin{tabular}{|c|c|}
  \hline
  Guide name &  Total transmission \\\hline
  TA W-1 open   &    0.986(6)(12)  \\\hline
  TA W-1 closed &    0.055(3)(12)  \\\hline
  TA W-2 open   &    0.973(4)(12)  \\\hline
  TA W-2 closed &    0.068(3)(12)   \\
  \hline
\end{tabular}
\end{center}
  \caption
	{Transmission values for the stainless steel UCN guides. 
	Open and closed refers to the UCN valve position.	
	The uncertainties given are statistical and systematical respectively.
	The total length of guide, valves and adapters are almost exactly 1\,m.
	}
 \label{stainless:trans}
\end{table}


\section{Summary}
\label{summary}

We have developed and used a prestorage method
to determine UCN transmission of tubular UCN guides.

Most importantly, the measurements 
provided a quality control for the UCN guides prior to
their installation at the PSI UCN source.

The results show excellent UCN transmission of the investigated
guides.
The measurements have shown that all guide
tubes have transmission values above 95$\%$ per meter.
The glass guides, which are dominantly used in the PSI UCN source, 
have transmissions  
above 98$\%$ per meter.

\section{Acknowledgements}
This work is part of the Ph.D. thesis of L.~G\"oltl.
We would like to thank all people which contributed to design and construction 
of our experiment.
A.~Anghel,
P.~Bucher, U.~Bugman, M.~M\"ahr and the AMI shop, 
F.~Burri,
M.~Dubs,
J.~Ehrat, 
M.~Horisberger,
R.~Knecht,
M.~Meier,
M.~M\"uller and his shop,
T.~Rauber,
P.~R\"uttimann, 
R.~Schelldorfer,
T.~Stapf,
J.~Welte (all PSI);
T.~Brenner (ILL);
F.~Lang, H.~H\"ase (S-DH);
J.~St\"adler (GlasForm Gossau);
M.~Klein (C-DT).
TU Munich (E18) allowed the use of their table installed at PF2.
Support by the Swiss National Science Foundation
Projects 200020\_137664 and 200020\_149813 is gratefully acknowledged.
\hbadness=10000


\end{document}